\documentstyle [12pt] {article}
\title{A SIMPLE DETERMINATION OF THE (LOGARITHMIC) CORRECTIONS
OF BLACK HOLE ENTROPY "WITHOUT KNOWING THE DETAILS OF QUANTUM
GRAVITY"}

\author{Vladan Pankovi\'c$^{\ast,\sharp}$,
Simo Ciganovi\'c$^\sharp$, Jovan Ivanovi\'c$^\sharp$\\
$^\ast$Department of Physics, Faculty of Sciences, 21000 Novi
Sad,\\ Trg Dositeja Obradovi\'ca 4. , Serbia, vdpan@neobee.net \\
$^\sharp$Gimnazija, 22320 Indjija, Trg Slobode 2a, Serbia \\}

\date {}
\begin{document}
\maketitle

 PACS number :   04.70.Dy

\begin {abstract}
In this work, starting by simple, approximate (quasi-classical)
methods presented in our previous works, we suggest a simple
determination of the (logarithmic) corrections of (Schwarzschild)
black hole entropy "without knowing the details of quantum
gravity"( Fursaev). Namely, in our previous works we demonstrated
that all well-known important thermodynamical characteristics of
the black hole (Bekenstein-Hawking entropy, Bekenstein
entropy/surface quantization and Hawking temperature) can be
effectively reproduced starting by simple supposition that black
hole horizon circumference holds integer number of reduced Compton
wave lengths corresponding to mass (energy) spectrum of a small
quantum system. (Obviously it is conceptually analogous to Bohr
quantization postulate interpreted by de Broglie relation in Old,
Bohr-Sommerfeld, quantum theory.) Especially, black hole entropy
can be presented as the quotient of the black hole mass and the
minimal mass of small quantum system in ground mass (energy)
state. Now, we suppose that black hole mass correction is simply
equivalent to negative classical potential energy of the
gravitational interaction between black hole and small quantum
system in ground mass (energy) state. As it is not hard to see
absolute value of the classical potential energy of gravitational
interaction is identical to black hole temperature. All this,
according to first thermodynamical law, implies that first order
entropy correction holds form of the logarithm of the surface with
coefficient -0.5. Our result, obtained practically
quasi-classically, "without knowing the details of quantum
gravity", is equivalent to result obtained by loop quantum gravity
and other quantum gravity methods for macroscopic black holes.
\end {abstract}
\vspace {1.5cm}

As it is well-known [1]-[11] there are different quantum gravity
theoretical methods for determination of the thermodynamical
characteristics, especially entropy and entropy correction, of the
black holes (loop quantum gravity with different sub-theories,
string theories with different sub-theories, two-dimensional
conformal symmetry near horizon, etc.). Surprisingly, practically
all these methods yield almost equivalent final results. Or, as it
has been pointed out by Carlip: "… today we suffer an
embarrassment of riches: many different approaches to quantum
gravity yield the same entropy, despite counting very different
states. This "universality" suggests that some underlying feature
of the classical theory may control the quantum density of
states." [1] Fursaev stated analogously: "In the last few years
there also appeared a hope that understanding of black hole
entropy may be possible even without knowing the details of
quantum gravity. The thermodynamics of black holes is a low energy
phenomenon, so only a few general features of the fundamental
theory may be really important." [10]

In this work, starting by simple, approximate (quasi-classical)
methods presented in our previous works [12]-[15], we shall
suggest a simple determination of the (logarithmic) corrections of
(Schwarzschild ) black hole entropy "without knowing the details
of quantum gravity"(Fursaev). Namely, in our previous works we
demonstrated that all well-known important thermodynamical
characteristics of the black hole (Bekenstein-Hawking entropy,
Bekenstein entropy/surface quantization and Hawking temperature)
can be effectively reproduced starting by simple supposition that
black hole horizon circumference holds integer numbers of reduced
Compton wave lengths corresponding to mass (energy) spectrum of a
small quantum system. (On the one hand, it is, obviously,
conceptually analogous to Bohr quantization postulate interpreted
by de Broglie relation in Old, Bohr-Sommerfeld, quantum theory. On
the other hand, our method can be considered as an extreme,
quasi-classical simplification and approximation of more accurate
quantum gravity methods, e.g. Copeland, Lahiri string theoretical
method [16], etc.) Especially, black hole entropy can be presented
as the quotient of the black hole mass and the minimal mass of
small quantum system in ground mass (energy) state. Now, we shall
suppose that black hole mass correction is simply equivalent to
negative classical potential energy of the gravitational
interaction between black hole and small quantum system. As it is
not hard to see absolute value of the classical potential energy
of gravitational interaction is identical to black hole
temperature. All this, according to first thermodynamical law,
implies that first order entropy correction holds form of the
logarithm of the surface with coefficient -0.5. Our result,
obtained practically quasi-classically, "without knowing the
details of quantum gravity", is equivalent to result obtained by
loop quantum gravity [2] and other quantum gravity methods for
macroscopic black holes.

Firstly, we shall repeat our previous results.

Consider a Schwarzschild black hole with mass $M$ and
corresponding horizon radius
\begin {equation}
     R=2M               .
\end {equation}

Suppose that, for "macroscopic" (with mass many time larger than
Planck mass, i.e. 1) black hole, at horizon surface there is a
small (with "microscopic" masses, i.e. masses smaller than Planck
mass, i.e. 1) quantum system.

Suppose the following condition
\begin {equation}
    m_{n} R = n \frac {1}{2\pi},
    \hspace{1cm} {\rm for}  \hspace{0.5cm} m_{n} \ll M  \hspace{0.5cm} {\rm and} \hspace{0.5cm}n=1,
      2,...
\end {equation}
where $m_{n}$ for  $\hspace{1cm} {\rm for}  \hspace{0.5cm} m_{n}
\ll M  \hspace{0.5cm} {\rm and} \hspace{0.5cm}n=1,
      2,... $
      represent the mass (energy) spectrum of
given quantum system. It corresponds to expression
\begin {equation}
      2\pi R = n \frac {1}{ m_{n}} = n \lambda_{rn}
      \hspace{1cm} {\rm for}  \hspace{0.5cm} m_{n} \ll M  \hspace{0.5cm} {\rm and} \hspace{0.5cm}n=1,
      2,...           .
\end {equation}
were $2\pi R$ represents the circumference of the black hole
horizon while
\begin {equation}
        \lambda_{rn}= \frac {1}{ m_{n}}
\end {equation}
represents $n$-th reduced Compton wavelength of mentioned small
quantum system with mass $ m_{n}$ for $n = 1, 2, …$ . Expression
(3) simply means that {\it circumference of the black hole horizon
holds exactly} $n$ {\it corresponding} $n$-{\it th reduced Compton
wave lengths of a quantum system with mass} $ m_{n}$ {\it captured
at the black hole horizon surface}, for $n = 1, 2, …$ . Obviously,
it is essentially analogous to well-known Bohr's angular momentum
quantization postulate interpreted via de Broglie relation.
However, there is a principal difference with respect to Bohr's
atomic model. Namely, in Bohr's atomic model different quantum
numbers $n = 1, 2, …$ , correspond to different circular orbits
(with circumferences proportional to $n^{2} = 1^{2},  2^{2}, …$).
Here any quantum number $n = 1, 2, …$ corresponds to the same
circular orbit (with circumference $2\pi R$).

According to (2) it follows
\begin {equation}
       m_{2} = n \frac {1}{2\pi R} = n \frac {1}{4\pi M} \equiv n m_{1}
       \hspace{1cm} {\rm for}  \hspace{0.5cm} m_{n} \ll M  \hspace{0.5cm} {\rm and} \hspace{0.5cm}n=1,
      2,...
\end {equation}
where
\begin {equation}
       m_{1} = \frac {1}{4\pi M}
\end {equation}
represents the minimal mass (ground energy level) of the small
quantum system. Obviously, $ m_{1}$ depends of $M$ so that $
m_{1}$ decreases when $M$ increases and vice versa. For a
"macroscopic" black hole, i.e. for $M \gg 1$ it follows $m_{1}\ll
1 \ll M$.

It is not hard to see that quotient of $M$ and $m_{1}$ represents
Bekenstein-Hawking black hole entropy, i.e.
\begin {equation}
       S = \frac {M}{m_{1}} = 4\pi M^{2} = \frac {A}{4}          ,
\end {equation}
where
\begin {equation}
       A = 4\pi R^{2}= 16\pi M^{2}= 4S
\end {equation}
represents the black hole surface area. Obviously, it represents
an interesting mechanical interpretation of the black hole
entropy.

Differentiation of (7) yields
\begin {equation}
      dS = 8\pi M dM
\end {equation}
or, after simple transformation,
\begin {equation}
      dM = \frac {1}{8\pi M } dS                     .
\end {equation}
Expression (10), representing the first thermodynamical law,
implies that term
\begin {equation}
      T = \frac {1}{8\pi M } = \frac {m_{1}}{2}
\end {equation}
represents Hawking black hole temperature. Moreover, expression
(11) points out that black hole temperature represents only one
half of the minimal mass of small quantum system. Since given
minimal mass represents difference between two neighboring mass
(energy) levels, it means that black hole temperature cannot lift
the small quantum system from some lower in the next, higher mass
(energy level).

Further, by approximate changing of the differentials by finite
differences, (9) turns out approximately, in
\begin {equation}
       \Delta S = 8\pi M \Delta M  \hspace{1cm}   {\rm for}   \Delta M \ll M .
\end {equation}
Now, assume
\begin {equation}
       \Delta m= n m_{1} \hspace{1cm}   {\rm for}  \hspace{1cm}  n=1,
      2,...
\end {equation}
which, according to (6), after substituting in (12), yields
\begin {equation}
       \Delta S = 2n  \hspace{1cm}   {\rm for}  \hspace{1cm}  n=1,
      2,...
\end {equation}
or, according to (8),
\begin {equation}
       \Delta A = (2n) 4 = (2n) 2^{2}  \hspace{1cm}   {\rm for}  \hspace{1cm}  n=1,
      2,...          .
\end {equation}
Obviously, expression (14) and (15) represent Bekenstein
quantization of the black hole entropy and horizon surface.

In this way we have reproduced, i.e. determined exactly in a
mathematically and physically simple way, three well-known most
important characteristics of Schwarzschild black hole
thermodynamics, Bekenstein-Hawking entropy (7), Hawking
temperature (11), and Bekenstein quantization of the entropy (14)
and horizon surface (15).

Now suppose that gravitational interaction between black hole and
small quantum systems implies effective increase of the black hole
mass. Then, quasi-classically corrected black hole mass, in the
simplest, first order, approximation can be, quite naturally,
according to (1), (6), and (11) presented by
\begin {equation}
  M_{cor} = M + \frac {Mm_{1}}{R} = M + \frac {1}{8\pi M }= M + T
\end {equation}
where $\frac {Mm_{1}}{R}$ represents the negative value of the
classical potential of gravitational interaction between black
hole and small quantum system in the ground mass (energy) state.
Obviously, absolute value of the classical potential of
gravitational interaction is equivalent to black hole temperature.
It represents an interesting mechanical interpretation of the
black hole temperature.

Differentiation of (16) yields
\begin {equation}
  dM_{cor} = dM - \frac {1}{8\pi M^{2} }dM            .
\end {equation}

Suppose, further, that given gravitational interaction, in the
first order approximation, does not change effectively the black
hole temperature even if it yields non-trivially corrected black
hole entropy $S_{cor}$.

Finally, suppose that for corrected black hole mass, entropy and
(trivially corrected) temperature satisfy first thermodynamical
law
\begin {equation}
   dM_{cor} = TdS_{cor}
\end {equation}
that, according to (11), (17), yields
\begin {equation}
   dS_{cor}= 8\pi M(dM - \frac {1}{8\pi M^{2} }) dM) = 8\pi MdM - \frac {1}{M}dM .
\end {equation}
It, after simple integration, yields,
\begin {equation}
  S_{cor}= 4\pi M^{2}- \ln [M] =  4\pi M^{2} - \frac {1}{2}\ln [\frac {16\pi M^{2}}{16\pi}]
\end {equation}
or, after simple transformations and according to (8),
\begin {equation}
 S_{cor}= 4\pi M^{2} - \frac {1}{2}\ln [16\pi M^{2}] + \frac {1}{2}\ln [16\pi] = \frac {A}{4} - \frac {1}{2} \ln [A]  + \frac {1}{2}ln [16\pi] .
\end {equation}

As it is not hard to see our result, obtained, obviously,
practically quasi-classically, "without knowing the details of
quantum gravity", excellently corresponds to result obtained by
loop quantum gravity [2] and other quantum gravity methods for
macroscopic black holes.

Finally, in conclusion, we can  repeat and point out the
following. In our previous works we demonstrated that all
well-known important thermodynamical characteristics of the black
hole (Bekenstein-Hawking entropy, Bekenstein entropy/surface
quantization and Hawking temperature) can be effectively
reproduced starting by simple supposition that black hole horizon
circumference holds integer numbers of reduced Compton wave
lengths corresponding to small quantum systems. Especially, black
hole entropy can be presented as the quotient of the black hole
mass and the minimal mass of small quantum system. Now, we
supposed that black hole mass correction is simply equivalent to
negative value of the classical potential energy of the
gravitational interaction between black hole and smallest quantum
system. As it is not hard to see absolute value of the classical
potential energy of gravitational interaction is identical to
black hole temperature. All this, according to first
thermodynamical law, implies that first order entropy correction
holds form of the logarithm of the surface with coefficient -0.5.
Our result, obtained practically quasi-classically, "without
knowing the details of quantum gravity", is equivalent to result
obtained by loop quantum gravity and other quantum gravity methods
for macroscopic black holes.

\section {References}

\begin {itemize}

\item [[1]] S. Carlip, {\it Black Hole Entropy and the Problem of Universality}, gr-qc/0702094
\item [[2]] A. Corichi, J. Diaz-Paolo, E. Fernadnez-Borja, {\it Loop Quantum Gravity and Planck-size black hole entropy}, gr-qc/0703116 and reference therein
\item [[3]] A. Ashtekar, J. Baez, A. Corichi, K. Krasnov, {\it Quantum Geometry and Black Hole Entropy}, gr-qc/9710007
\item [[4]] K. A.Meissner, {\it Black Hole Entropy in Loop Quantum Gravity}, gr-qc/0407052
\item [[5]] A. Strominger, C. Vafa, {\it Microscopic Origin of the Bekenstein-Hawking Entropy}, hep-th/9601029
\item [[6]] G. Amelino-Camelia, M. Arzano, G. Mandanizi, {\it Black-hole Thermodynamics with Modified Dispersion Relations and Generalized Uncertainty Principles}, gr-qc/0506110
\item [[7]] S. Hod, {\it Higher-order Corrections to the Entropy and Area of Black Holes}, hep-th/0405235
\item [[8]] S. Kloster, J. Brannlund, A. DeBenedictis, {\it Phase-space and Black Hole Entropy of Higher Genus Horizons in Loop Quantum Gravity}, gr-qc/0702036
\item [[9]] M. Arzano, {\it Black Hole Entropy, log Corrections and Quantum Ergosphere}, gr-qc/0512071
\item [[10]] D. V. Fursaev, {\it Can One Understand Black Hole Entropy without Knowing Much about Quantum Gravity?}, gr-qc/0404038
\item [[11]] D. N. Page, {\it Black Holes with Less Entropy than A/4}, gr-qc/0109040
\item [[12]] V.Pankovic, M.Predojevic, P.Grujic, {\it A Bohr's Semiclassical Model of the Black Hole Thermodynamics},
gr-qc/0709.1812
\item [[13]]  V. Pankovic, J. Ivanovic, M.Predojevic, A. M. Radakovic,
{\it The Simplest Determination of the Thermodynamical
Characteristics of Schwarzschild, Kerr and Reisner-Nordstr$\ddot
{o}$m black hole}, gr-qc/0803.0620
\item [[14]] V. Pankovic, S. Ciganovic, R. Glavatovic, {\it The Simplest Determination of the Thermodynamical Characteristics of Kerr-Newman Black Hole}, gr-qc/0804.2327
\item [[15]] V. Pankovic, R. Glavatovic, S. Ciganovic, D-H. Petkovic, L. Martinovic, {\it Single Horizon Black Hole "Laser" and a Solution of the Information Loss Paradox}, gr-qc/0807.1840
\item [[16]] E. J. Copeland, A.Lahiri, Class. Quant. Grav. , {\bf 12} (1995) L113 ; gr-qc/9508031

\end {itemize}

\end {document}